\documentclass[aps,prl]{revtex4}
\usepackage{amsmath}
\usepackage{epsfig,epsf}
\usepackage{graphicx}
\usepackage{verbatim}
\usepackage{epstopdf}

\def\beq{\begin{equation}}
\def\eeq{\end{equation}}
\def\bea{\begin{eqnarray}}
\def\eea{\end{eqnarray}}
\def\eq#1{{Eq.~(\ref{#1})}}
\def\fig#1{{Fig.~\ref{#1}}}

\newcommand{\bas}{\bar{\alpha}_s}
\newcommand{\as}{\alpha_s}
\newcommand{\Lb}{\left(}
\newcommand{\Rb}{\right)}
\newcommand{\nn}{\nonumber}

\begin{document}


\voffset1.5cm
\title{Nuclear modification factor for gluon  jets }
\author{Eugene Levin}
\affiliation{
Department of Particle Physics, School of Physics and Astronomy,
Raymond and Beverly Sackler
 Faculty of Exact Science,Tel Aviv University, Tel Aviv, 69978, Israel\\ 
and\\
Departamento de F\'\i sica, Universidad T\'ecnica Federico Santa Maria
 and Centro 
Cient��fico-Tecnol$\acute{o}$gico de Valpara\'\i so,Casilla 110-V, 
Valparaiso, Chile}
 
\date{\today}
\begin{abstract}
It is shown that the nuclear modification factor  can be smaller that unity for jet production at small $x$ and  at large transverse  momentum without any violation of the factorization theorem and the initial state effects is able to explain the nuclear modification factor of the order of the one measured by ATLAS. In other word, initial state effects are able to describe the jet quenching for the gluon jet production.
\end{abstract}
\maketitle

\section{Introduction}

\begin{figure}[ht]
\begin{tabular}{c c}
\includegraphics[width=6cm]{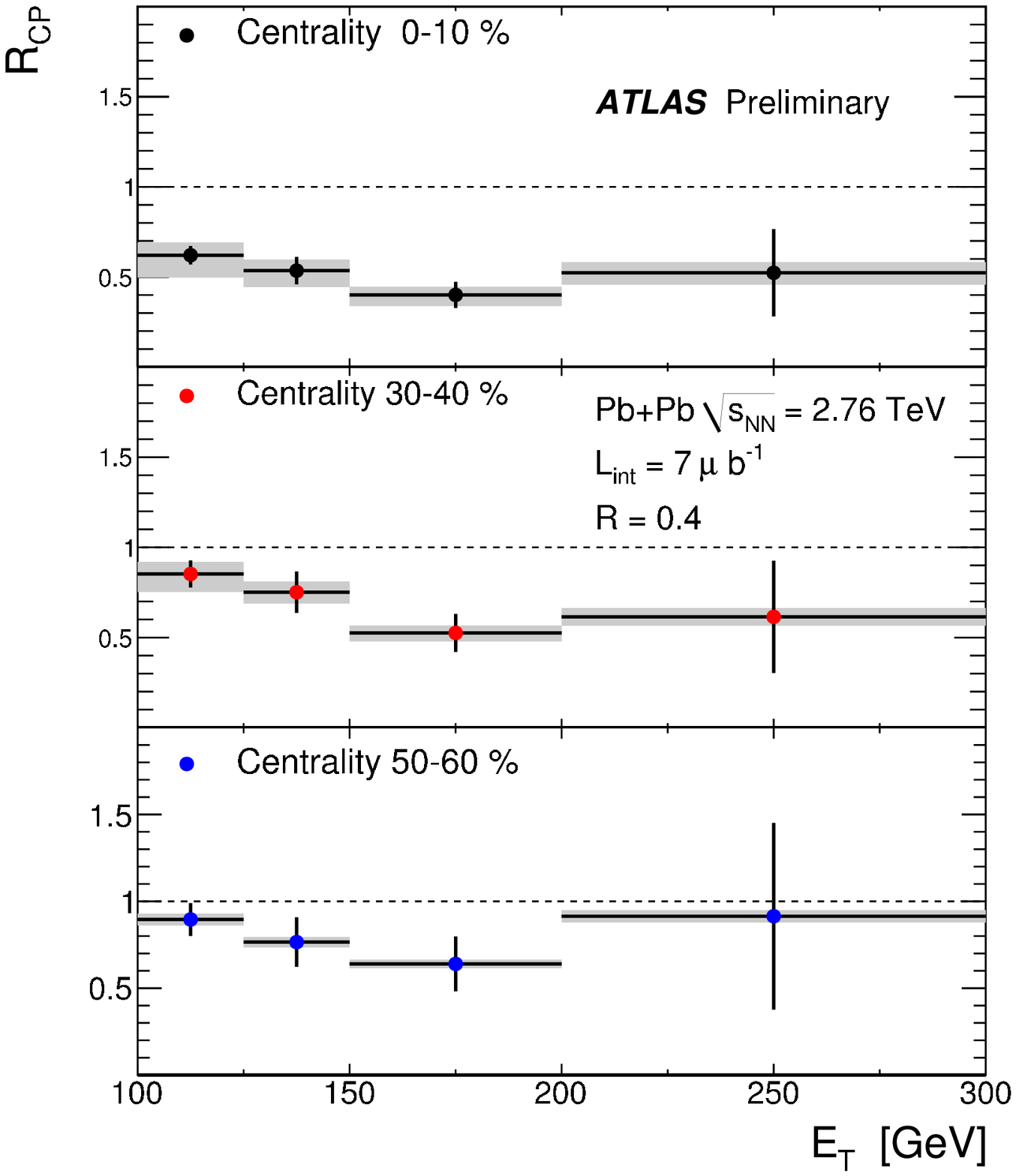}& \includegraphics[width=6cm]{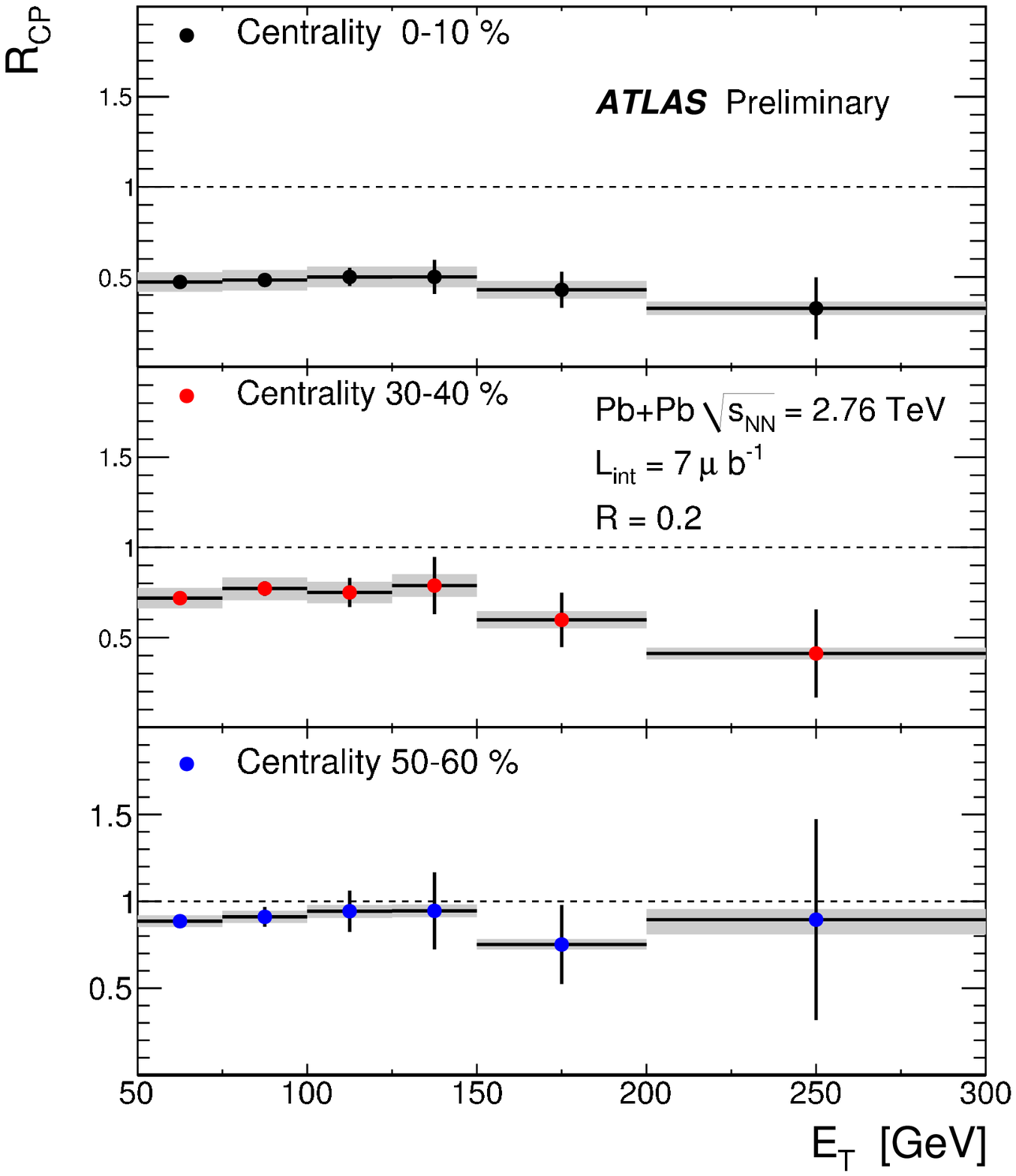}\\
\fig{nmf}-a & \fig{nmf}-b\\

\end{tabular}
\caption{ Nuclear modification factor(NMF) for jet production.  The figures are taken from Ref.\cite{ATLAS}
}
\label{nmf}
\end{figure}

Recently, the ATLAS collaboration  \cite{ATLAS} has measured the NMF factor for jet production.  It turns out that the
 NMF does not depend on the size of the cone, in which hadrons from the  jet decay were measured, and its value is considerably smaller than 1 (see \fig{nmf}). From \fig{nmf} one can see that the suppression is larger  for the events with small centrality. Such an independence on the size of the come allows us to assume that all hadrons from the jet decay were measured in the ATLAS experiment.  If it is so,
  these  data contradict the QCD factorization theorem
 \cite{FT1,FT2,FT3,FT4} (see discussion below) which is one the most solid result of QCD. 
 
 These data encourage me to ask two theoretical questions.  The  first question: could we obtain the NMF less that unity for production of gluon  at very high energy (small $x$)  and with large value of transverse momentum, without violation of the factorization theorem? The second one: can be the value of NMF of the order of the value measured by ATLAS?  We would like to emphasize that the goal of this letter to answer these two theoretical questions but not to describe the  ATLAS experimental data.

Let us first discuss the factorization theorem and why this theorem at first sight leads to the NMF equal to 1.

Indeed, the factorization theorem states that the  inclusive cross section for gluon production with rapidity $Y'$
and transverse momentum $p_T$ can be written as\cite{FT1,FT2,FT3,FT4}
\beq \label{FT}
\frac{d^2 \sigma \Lb Y',p_T\Rb}{d Y' d^2 p_T}\,=\,\sigma_{hard} \bigotimes x_1 G_{Pr}\Lb x_1,\mu_F\Rb
x_2 G_{Tr}\Lb x_2,\mu_F\Rb\,\,\,\,\,\mbox{where}\,\,\,\,x_1\,=\frac{p_T}{\sqrt{s}}e^{Y'}\,\,\mbox{and}\,\,x_2\,=\frac{p_T}{\sqrt{s}}e^{- Y'} \eeq
where $\sqrt{s}= W$ is the energy of colliding particles and $\bigotimes$ denotes  all necessary  integrations while  $\mu_T$ is the factorization scale which should be chosen of the order of $p_T$ of produced gluon. \eq{FT} has a  simple meaning, namely that the production of a
high $p_T$ gluon is proportional to the  cross section of the interaction of two partons, with fractions of energy $x_1$ and $x_2$, multiplied by the probability
 to find them in the projectile ($G_{Pr}$) and target ($G_{Tr}$).  All interactions in the final state as well as  corrections to the initial state (for, example, due to shadowing) lead to small small contributions of the order of $1/p^2_T$\footnote{ It should be mentioned that the final state interaction for production of the gluon jet with large $p_T$ is suppressed even for $k_T$-factorization\cite{SKTF} that has less solid theoretical basis but  which we will use below for estimates of the size of the NMF.}.
 
   At short distances (large $p_T$) the gluon structure function for a nucleus
is equal to  
\beq \label{AG}
 xG_A\Lb x, \mu_T\Rb \,=\,A xG_P\Lb x, \mu_T\Rb
\eeq
 where $G_P$ is the nucleon gluon structure function. Therefore, 
\beq \label{FT1}
\frac{d^2 \sigma \Lb A_1 + A_2|Y',p_T\Rb}{d Y' d^2 p_T}\,\,=\,\,\, A_1 A_2 \frac{d^2 \sigma \Lb P + P |Y',p_T\Rb}{d Y' d^2 p_T} 
\eeq
where $ d^2 \sigma \Lb P + P |Y',p_T\Rb \Big{/}d Y' d^2 p_T$ is the cross section of inclusive production in proton-proton collisions.

In other words the nuclear modification factor (NMF) $R_{AA}$, which is defined as

\beq \label{FT2}
R_{AA}\,\,\,=\,\,\,\frac{\frac{d^2 \sigma \Lb A_1 + A_2|Y',p_T\Rb}{d Y' d^2 p_T}}{ N_{coll}\,
\frac{d^2 \sigma \Lb P + P |Y',p_T\Rb}{d Y' d^2 p_T}  }\,\,\,=\,\,\,1
\eeq

where $N_{coll}$ is the number of collision that is equal to $A_1 \,A_2$, in the case of large $p_T$.

The main result of this letter is that \eq{AG} should  to be replaced by
\beq \label{AGF}
 xG_A\Lb x, \mu^2_F/Q^2_s\Lb A; x_0\Rb\Rb \,=\,A xG_P\Lb x, \mu^2_F/Q^2_s\Lb A ; x_0\Rb\Rb\,=A xG_P\Lb x, \mu^2_F/\Big( A^{1/3}Q^2_s\Lb P ; x_0\Rb\Big)\Rb \, \,\,\mbox{for}\,\,\mu_F \, \approx \,p_T \gg\,\,Q_{s}\Lb A; x\Rb
\eeq
where $Q_s\Lb A; x\Rb$  ($ Q_s\Lb P; x\Rb$) are  the saturation momenta at $Y = \ln (1/x)$ for the nucleus and the proton, respectively.  
$Y_0 = \ln(1/x_0)$ is the rapidity at which  the evolution starts. In other word, we state that the linear DGLAP or BFKL evolution should be started at different initial condition at low energy ($Y=Y_0$).

It follows from \eq{AGF},  that \eq{FT2} reads as 
\beq \label{FT3}
R_{AA}\,\,\,=\,\,\,\frac{\frac{d^2 \sigma \Lb A_1 + A_2|Y',p_T\Rb}{d Y' d^2 p_T}}{ N_{coll}\,
\frac{d^2 \sigma \Lb P + P |Y',p_T\Rb}{d Y' d^2 p_T}  }\,\,\,= \,\,\,\frac{x_1 G_P\Lb x_1, \mu^2_F/Q^2_s\Lb A; x_{10}\Rb \Rb\, x_2 G_P\Lb x_2, \mu^2_F/Q^2_s\Lb A; x_{20}\Rb\Rb}{x_1 G_P\Lb x_1, \mu^2_F/Q^2_s\Lb P; x_{10}\Rb \Rb x_1 G_P\Lb x_2, \mu^2_F/Q^2_s\Lb P; x_{20}\Rb \Rb}
\eeq

For the sake of numerical estimates 
we will derive the  QCD factorization formula from the $k_T$-factorization, in order to specify the factorization scale $\mu_F$ and  all numerical coefficients in \eq{FT3}.
Using  $k_T$ factorization \cite{KTF1,KTF2,KTF3,KTF4}, the inclusive cross section has the form

\beq \label{F1}
\frac{d \sigma}{d y \,d^2 p_{T}}\,\,=\,\,\frac{2 \pi \bas}{p^2_T}\int d^2 k_T\,\,\phi^{A_1}_G\Lb x_1;\vec{k}_T\Rb\,\phi^{A_2}_G\Lb x_2;\vec{p}_T -\vec{k}_T\Rb
\eeq
where $\phi^{A_i}_G$ is the probability to find a gluon in the nucleus ($A_i$),  that carries the fraction of energy  $x_i$  with $k_T$ transverse momentum,
 and $ \bas \,= \,\as N_c/\pi$, where $N_c$ is equal to the number of colours in the $SU_{N_c}$ colour group.
At large values of $p_T$ we can re-write \eq{F1} in the form
\bea
\frac{d \sigma}{d y \,d^2 p_{T}}\,\,&=&\,\,\frac{2 \pi \bas}{p^2_T}\Big\{ \int^{p_T} d^2 k_T\phi^{A_1}_G\Lb x_1;\vec{k}_T\Rb\phi^{A_2}_G\Lb x_2;\vec{p}_T\Rb\,+\,\phi^{A_1}_G\Lb x_1;\vec{p}_T\Rb\int^{p_T} d^2 k_T\phi^{A_2}_G\Lb x_2;\vec{k}_T\Rb\Big\}\,\,\nn\\
&=&\,\,\frac{2 \pi \bas}{p^2_T}\,\frac{d}{d p^2_T}\Big( 
x_1 G_{A_1}\Lb x_1,p^2_T\Rb
x_2 G_{A_2}\Lb x_2,p^2_T\Rb\Big)\label{F2}
\eea

For the purpose of deriving \eq{F2}, we used the relation between the un-integrated gluon density $\phi$ and the gluon structure function (see for example Ref.\cite{GLR}):

\beq \label{PHIG}
xG\Lb x, Q^2\Rb \,\,=\,\,\int^{Q^2}_0\,\phi\Lb k^2_T\Rb d k^2_T
\eeq 
Using \eq{F2} one can see that \eq{FT3} can be rewritten as
\beq \label{FF}
R_{AA}\,\,\,=\,\,\frac{\frac{d}{d p^2_T}\Big( x_1 G_P\Lb x_1, p^2_T/Q^2_s\Lb A; x_{10}\Rb \Rb\, x_2 G_P\Lb x_2, p^2_T/Q^2_s\Lb A; x_{20}\Rb\Rb\Big)}{\frac{d}{d p^2_T}\Big(x_1 G_P\Lb x_1, p^2_T/Q^2_s\Lb P; x_{10}\Rb \Rb\, x_2 G_P\Lb x_2, p^2_T/Q^2_s\Lb P; x_{20}\Rb\Rb\Big)}
\eeq
\begin{boldmath}
\section{ Relation  $ xG_A\Lb x, p^2_T/Q^2_s\Lb A; x_0\Rb\Rb \,=\,A xG_P\Lb x, p^2_T/\Big(A^{1/3}Q^2_s\Lb P ; x_0\Rb\Big)\Rb$}
\end{boldmath}
The proof of \eq{AGF} is based on two observations. First, the equations in the region of low $x$ can be written
 as an evolution in rapidity, both in the  approach of DGLAP \cite{DGLAP}
 and in the BFKL \cite{BFKL} approach. They both  have the following general form
\beq \label{EVEQ}
\frac{\partial xG\Lb Y; p^2_T\Rb}{\partial Y}\,
\,=\,\,\int d^2 k_T \,K\Lb p_T, k_T\Rb\, xG\Lb Y, k^2_T\Rb
\eeq

 where $Y  =  \ln(1/x)$.

Second, since  for large $p_t$ we can restrict  ourselves to the linear evolution:  DGLAP or BFKL, one can see that the solution of \eq{EVEQ} can be written as
\beq \label{SOL1}
xG\Big( Y, \xi\equiv \,\ln\Lb p^2_T/\Lambda^2_{QCD}\Rb\Big)\,\,=\,\,\int d \xi' \,G\Lb Y - Y_0, \xi - \xi'\Rb \,xG_{in}\Lb Y=Y_0, \xi'\Rb
\eeq
where $G\Lb Y - Y_0,\xi - \xi'\Rb$ is the Green function of \eq{EVEQ}: the solution to \eq{EVEQ} with the initial condition
 $G\Lb Y=Y_0,\xi - \xi'\Rb \,=\,\delta\Lb \xi - \xi'\Rb$. $ xG_{in}\Lb Y=Y_0, \xi'\Rb$ is the initial condition for the gluon structure function.

 For linear evolution the solution to the equation reveals the following property of the propagator \cite{GLR,MUSH,KLM}:
\beq \label{UN}
xG\Big( Y, \xi\equiv \,\ln\Lb p^2_T/\Lambda^2_{QCD}\Rb \Big) \,\,=\,\,\int\, d \xi' \,
\,G\Lb Y - Y' , \xi -\xi'\Rb\,xG\Lb  Y' , \xi'\Rb
\eeq
for any value of $Y'$.  We can prove \eq{UN} by solving \eq{EVEQ} using the double Mellin transform
\beq \label{METR}
xG\Lb Y, \xi\Rb\,\,\,=\,\,\int^{\epsilon + i \infty}_{\epsilon - i \infty}  \frac{d \omega}{2 \pi i}\,\int^{\epsilon + i \infty}_{\epsilon - i \infty}  \frac{d \gamma}{2 \pi i}\,e^{ \omega Y +  \gamma \xi}\,g\Lb \omega,\gamma\Rb
\eeq

Plugging  \eq{METR} into \eq{EVEQ} we obtain the solution 
\beq \label{SOL}
\omega \,\,=\,\,\omega\Lb \gamma\Rb
\eeq
where $\omega\Lb \gamma \Rb$ is the Mellin image of the kernel in \eq{EVEQ}.  It can be verified  that the Green function has the form
\beq \label{GRF}
G\Lb Y - Y' , \xi\Rb\,\,\,=\,\,\,\int^{\epsilon + i \infty}_{\epsilon - i \infty}  \frac{d \gamma}{2 \pi i}\,e^{ \omega\Lb \gamma\Rb\Lb Y - Y'\Rb \,+\,\gamma \Lb \xi - \xi' \Rb} \,
\eeq
while the solution of \eq{SOL1} can be written in two equivalent forms
\beq \label{SOL2}
xG\Lb Y, \xi\Rb\,\,\,=\,\,\,\int^{\epsilon + i \infty}_{\epsilon - i \infty}  \frac{d \gamma}{2 \pi i}\,e^{ \omega\Lb \gamma \Rb Y +  \gamma \xi}\,g_{in}\Lb \gamma\Rb\,\,=\,\,\int^{\epsilon + i \infty}_{\epsilon - i \infty}  \frac{d \omega}{2 \pi i}\,e^{ \omega Y +  \gamma \Lb \gamma\Rb\,\xi}\,g_{in}\Lb \omega\Rb\eeq
where $g_{in} $ is the Mellin image of  $xG_{in}\Lb Y=Y_0, \xi'\Rb$ in \eq{SOL1}  and $\gamma \Lb \omega \Rb$ is the anomalous dimension which can be obtained as solution of \eq{SOL}.

Using \eq{GRF} and \eq{SOL2} for the r.h.s. of \eq{UN}, and integrating over $\xi'$ we obtain the l.h.s. of this equation.

Our standard approach to low $x$ evolution  consists of two steps. First,  we assume that there exists a small $x = x_0$ ($Y=Y_0=\ln(1/x_0)$) which is large
  ($ Y_0 \gg 1$), but at the same time small enough such that $\bas Y_0 \ll 1$ .
For this type of $Y_0$ we have
the theoretical formula     (the McLerram-Venugopalan formula \cite{MV} )  for the scattering amplitude of the dipole,which looks as follows in the most simplified form
\beq \label{MVF}
N\Lb r, b;Y_0\Rb\,\,=\,\, 1\,
\,-\,\,\exp\Lb - r^2Q^2_s\Lb Y_0,b\Rb/4\Rb
\eeq
where $N$ is the imaginary part of the scattering amplitude; $Q_s$ is the saturation scale at $Y=Y_0$ , $r$ is the dipole size and $b$ is the impact parameter.
The main features of  \eq{MVF} are that the amplitude of \eq{MVF} has the geometric scaling behaviour \cite{GS} at low energy ($Y=Y_0$), since it is a function of only one argument
  $N\Lb r, b;Y_0\Rb \,=\, N\Lb\tau \equiv r^2 Q_s\Lb Y_0,b\Rb \Rb$.
  
  The second step is to use the  evolution in the region $Y > Y_0$. In our case, when the typical size of the dipole is small ($r \approx 1/p_T$),
  we can use the linear BFKL or DGLAP equations. It should be stressed that we use the rigorous theoretical formula of \eq{MVF}, which shows the geometric scale behaviour only for the initial condition, for the linear evolution.

We will show below that for the initial condition $x_0G\Lb Y_0,k_T\Rb \,=\,S\,Q^2_s(Y_0) g\Lb \tau = k^2_T/Q^2_s\Lb Y\Rb\Rb$.  Using \eq{UN} with $Y'=Y_0$ we have
\bea \label{UN1}
xG\Big( Y, \xi\equiv \,\ln\Lb p^2_T/\Lambda^2_{QCD}\Rb \Big) \,\,&=&\,\,\int\, d \xi'\,
\,G\Lb Y - Y_0 , \xi -\xi'\Rb\,x_0G\Lb  Y_0 , \xi'\Rb\,\,\nn\\
 &=&\,\,S\, Q^2_s\Lb Y_0\Rb \int \frac{d k^2_T}{k^2_T}  G\Lb Y - Y_0 , p^2_T/k^2_T\Rb\,g\Lb  k^2_T/Q^2_s\Lb Y_0\Rb\Rb \nn\\
&=& \,\,\,S\, Q^2_s(Y_0)\,\int^{\epsilon + i \infty}_{\epsilon - i \infty}  \frac{d \gamma}{2 \pi i}\,e^{ \omega\Lb \gamma\Rb\Lb Y - Y_0\Rb}  \,\Lb \frac{p_T^2}{k^2_T}\Rb^\gamma \,g\Lb \frac{k^2_T}{Q^2_s\Lb Y_0\Rb}\Rb\nn\\
&=&\,\,S\, Q^2_s(Y_0)\,\int^{\epsilon + i \infty}_{\epsilon - i \infty}  \frac{d \gamma}{2 \pi i}\,e^{ \omega\Lb \gamma\Rb\Lb Y - Y_0\Rb}  \,\Lb \frac{p_t^2}{Q^2_s\Lb Y_0\Rb}\Rb^\gamma\int \frac{d \tau}{\tau}
\tau^{-\gamma}\,g\Lb \tau\Rb\nn\\
 &=&\,\,\,S\, Q^2_s(Y_0)\,\int^{\epsilon + i \infty}_{\epsilon - i \infty}  \frac{d \gamma}{2 \pi i}\,e^{ \omega\Lb \gamma\Rb\Lb Y - Y_0\Rb}  \,\Lb \frac{p_t^2}{Q^2_s\Lb Y_0\Rb}\Rb^\gamma f\Lb \gamma\Rb
\eea
where $\tau = k^2_T/Q^2_s$ and $S$ is the area of the target.
  In \eq{UN1} we used \eq{GRF}. The  function $f(\gamma)$ does not depend  on the character of the target, and it is the same for the nucleus and for the nucleon
 if we consider
\eq{MVF} for both. Since $Q_s\Lb A; Y_0\Rb\,\,=\,\,A^{1/3}\,Q_s\Lb P;Y_0\Rb$ and $S_A = A^{2/3}$ , one can see that \eq{AGF} follows from \eq{UN1}.

Now let us show that $x_0G\Lb Y_0,k_T\Rb \,=\,S\,Q^2_s(Y_0) g\Lb \tau \Rb$.
The relation 
 between $\phi^{A}_G\Lb x;\vec{k}_T\Rb$ and the dipole scattering amplitude $N\Lb x, r; b \Rb$ has the form \cite{KTINC}
\beq \label{PHN}
\phi^{A}_G\Lb x_0;\vec{k}_T\Rb\,\,=\,\,\frac{1}{\bas\,4 \pi}\,\int d^2 b \,d^2 r
e^{i \vec{k}_T\cdot \vec{r}}\,\,\nabla^2_\perp\,N^{A}_G\Lb Y_0 ; r; b \Rb
\eeq
where 
\beq \label{NG}
N^{A}_G\Lb Y_0; r,  b \Rb\,\,=\,\,2 \,N\Lb Y_0; r, b \Rb\,\,\,-\,\,\,N^2\Lb Y_0;r, b \Rb\xrightarrow{\eq{MVF}}\,
1\,\,-\,\,\exp\Lb - r^2 Q^2_s\Lb Y_0;b\Rb/2\Rb
\eeq
Plugging \eq{NG} into \eq{PHN} and  using the simplest model for the $b$ dependence, viz. $Q_s^2\Lb Y_0; b\Rb \,\,=\,\,Q_s\Lb Y_o\Rb \Theta\Lb R_A - b\Rb$ we obtain
\beq  \label{PHIN}
\phi^{A}_G\Lb x_0;\vec{k}_T\Rb\,\,=\phi^A_G\Lb \tau = k^2_T/Q^2_s\Lb Y_0\Rb\Rb\,\,=\,\,\,2 \pi \,S\, \tau \,\exp\Lb- 2 \,\tau\Rb
\eeq
Using \eq{PHIG} we calculate the gluon structure function at $Y=Y_0$. It is equal to
\beq \label{GIN}
x_0 G\Lb x_0, p^2_T\Rb \,\,=\,\,\frac{\pi}{2} S\,Q^2\Lb Y_)\Rb\,\Big( 1\,-\,\Lb 1 + 2\,\tau\Rb \,\exp\Lb - 2 \tau\Rb\Big)\,\,=\,\,S\,Q^2_s\Lb Y_0\Rb \,g \Lb \tau\Rb
\eeq
Therefore, we arrive at the result $x_0G\Lb Y_0,k_T\Rb \,=\,S\,Q^2_s(Y_0) g\Lb \tau \Rb$ with the particular function $g\Lb \tau\Rb$.  The corresponding function
 $f\Lb \tau\Rb$ (see \eq{UN1}) turns out to be equal to
$f\Lb \tau\Rb\,=\, 2^\gamma \Gamma\Lb 2 - \gamma\Rb/\gamma$.

\begin{figure}[h]
\includegraphics[width=8cm]{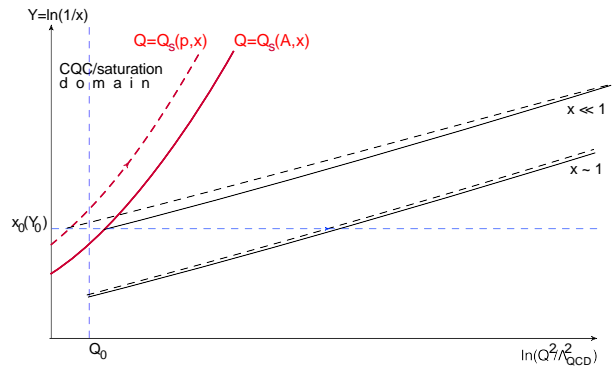}
\caption{ The semi-classical trajectories for the nucleus (solid lines) and nucleon (dashed lines)  gluon structure function in the region of large and small $x$. In the  Color Glass Condensate (CGC)/saturation domain the structure function has a geometric scaling behaviour.
}
\label{traj}
\end{figure}

\eq{UN1} is the main result of the paper. It should be stressed that solution which we found, does not lead to the geometric scaling behaviour since it depends on both variable: $x$ and $p_T$. However,  \eq{UN1} shows that  $p_T$ in the DGLAP evolution  scales with different momentum that depends on the target. In this sense this equation shows the $p_T$ scaling behaviour. The fact that in DGLAP evolution $p_T$ is scaled by the typical momentum in the initial condition is well known and follows directly from the conformal symmetry of the DGLAP equation at fixed QCD coupling. Two ideas are new in the proof of \eq{UN1}:  the evolution at low $x$ has to start with rather low initial $x_0$ and the initial condition at this $x = x_0$ has to show the geometric scaling behaviour. Both of these results   follows from the Color Glass Condensate/saturation approach and are rigorous results for QCD.

To illustrate the situation we consider the semi-classical approach to the solution of the DGLAP equations\cite{SCL}. In this approach we take the integral over $\gamma$ in \eq{UN1} using the steepest decent method and the solution is characterized by the trajectory: the line in $( \ln (1/x), \ln\Lb Q^2/\Lambda^2_{QCD}\Rb$ plane. The trajectory shows what values of $x_0$ and $Q_0$ in the initial condition are essential  to find the solution in the point $\Lb \ln (1/x), \ln \Lb Q^2/\Lambda^2_{QCD}\Rb\Rb$. \fig{traj} shows the trajectories for gluon structure function in the region of low $x$ and in the region of $x \sim 1$. These trajectories are denoted by solid and dashed lines for nucleus and nucleon targets, respectively. On can see that for $x \sim 1$ both trajectory for nucleus and for proton targets started at the value of $Q=Q_0$ which corresponds to the same scaling of $p_T$ for interaction with nucleus and nucleon. However, for the region of low $x$ the situation changes crucially and the evolution starts from the value of $Q_0$ that should be determined  by the initial condition given by the Color GlassCondensate/saturation approach. \fig{traj} illustrates the main result of the CGC/saturation approach: we cannot use the initial condition at fixed arbitrary $Q=Q_0$ since its contradicts the unitarity constraints.

~

\section{Estimates for  NMF}

\begin{figure}[ht]{
\begin{tabular}{c c c}
\includegraphics[width=7cm]{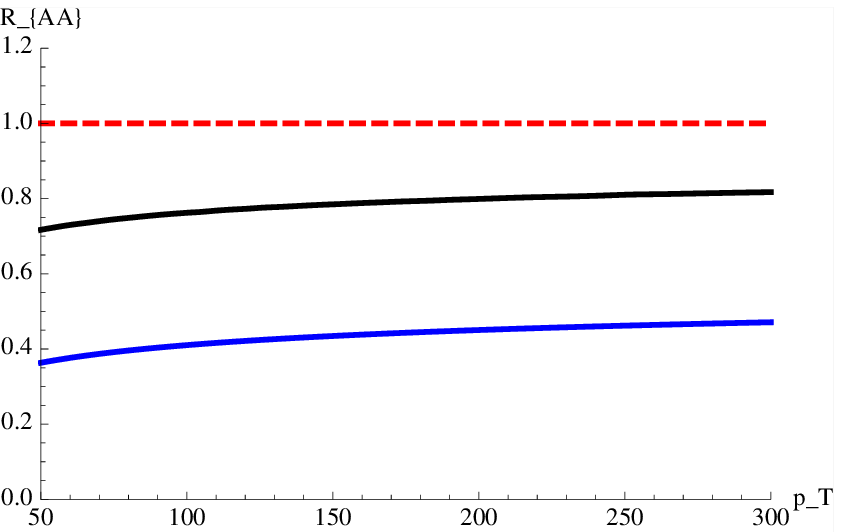} &~~~~~~~~~~~~&\includegraphics[width=7cm]{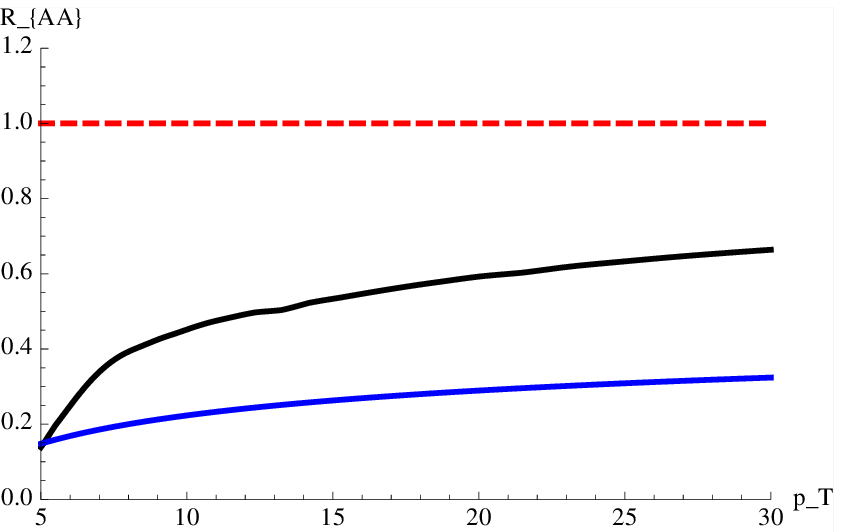}\\
\fig{raa}-a & & \fig{raa}-b\\
\end{tabular}
\caption{Nuclear modification factor (NMF) for jet production.  The black line shows $R_{AA}$ from \eq{FT3}
for MRSTW  parameterizations of the gluon structure functions (see Ref.\cite{MRSTW}. $R_{AA}$  from \eq{GFIN} is plotted by the blue line.  $\bas$ is taken to be $\bas=0.2$ and $x_1 = x_2 = 10^{-3}$
}
\label{raa}
}
\end{figure}

At first sight the  value for $R_{AA} $ could be calculated in a direct way, just by using \eq{FT3} and various different 
fits to the available DIS data, using the DGLAP evolution equations.  By plugging in \eq{FT3} the gluon structure functions 
that are available on the market (see Ref.\cite{DISSTR}) and 
 using $Q^2_s\Lb A; Y_0\Rb = A^{1/3} Q^2_s\Lb P; Y_0\Rb$ , then we obtain $R_{AA}$ for the gluon jet production at the LHC, in the central rapidity region for the lead-lead
 collisions. This estimates  are   shown by the black line in \fig{raa}. In \fig{raa} are plotted the NMF using the MRSTW parameterization \cite{MRSTW} but we check that the other parameterizations  give approximately
 the same value for $R_{AA}$.

However, this type of approach suffers from two major defects. First, in all parameterizations the running QCG coupling is used. 
 The transverse momenta $p_T$  enters to the expression for  the running QCD coupling as $\ln\Lb p^2_T/\Lambda_{QCD}\Rb $ and, therefore, they  are not  rescaled with
the saturation momenta. Second, the gluon structure functions, that we used,  where extracted from the experimental data using  the DGLAP evolution in transverse momenta, but not in $Y$ .

We believe that we can obtain more reliable estimates by using \eq{UN1}, \eq{SOL2} and \eq{GIN}. 
It is known that the leading order DGLAP anomalous dimension can be  approximated with an accuracy of 5$\%$ by the following expression \cite{EKL}

\beq \label{ANDI}
\gamma\,\,= \,\ \bas \Big( \frac{1}{\omega}\,-\,1\Big) \,\,\,\,\,\mbox{or}\,\,\,\,\,\,\omega\Lb \gamma\Rb \,\,=\,\,\frac{\as}{\gamma \,+\,\bas}
\eeq
Substituting $\omega(\gamma)$ from \eq{ANDI} and taking the integral using the steepest decent method,  we obtain the following answer:
\beq \label{GFIN}
xG\Lb Y, \tau= \Rb\,\,=\,\,f\Lb \gamma_{SP}\Rb\,\exp\Big( 2\, \sqrt{ \bas \,Y\,\ln \Lb  \tau\Rb} \,\,-\,\,\bas\,\ln\Lb \tau\Rb\Big)\,\,\,\,\mbox{with}\,\,\,\,\,\gamma_{SP}\,\,=\,\,\sqrt{\frac{\bas Y}{\ln \Lb \tau\Rb}}\,\,-\,\,\bas
\eeq
where $\tau= p^2_T/Q^2_s\Lb Y_0\Rb$ and $Y = \ln(1/x)$.

The blue curve in \fig{raa} describes $R_{AA}$ from
 \eq{FF} with the gluon structure function given by \eq{GFIN}. This estimate leads to smaller values of $R_{AA}= 0.4 \div 0.5$ (see \fig{raa}).

\begin{figure}
\begin{tabular}{c c c }
\includegraphics[width=5cm]{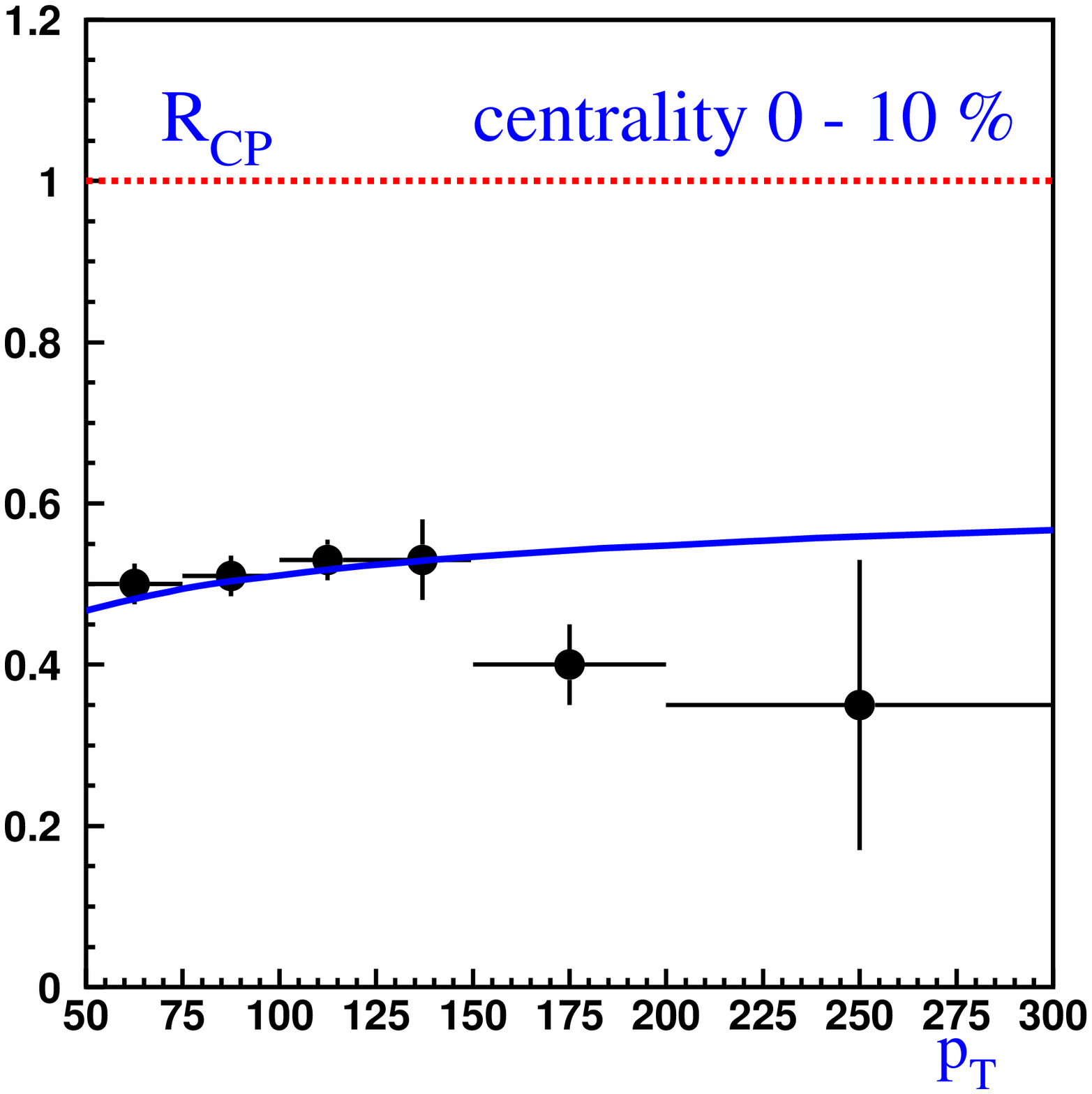}& \includegraphics[width=5cm]{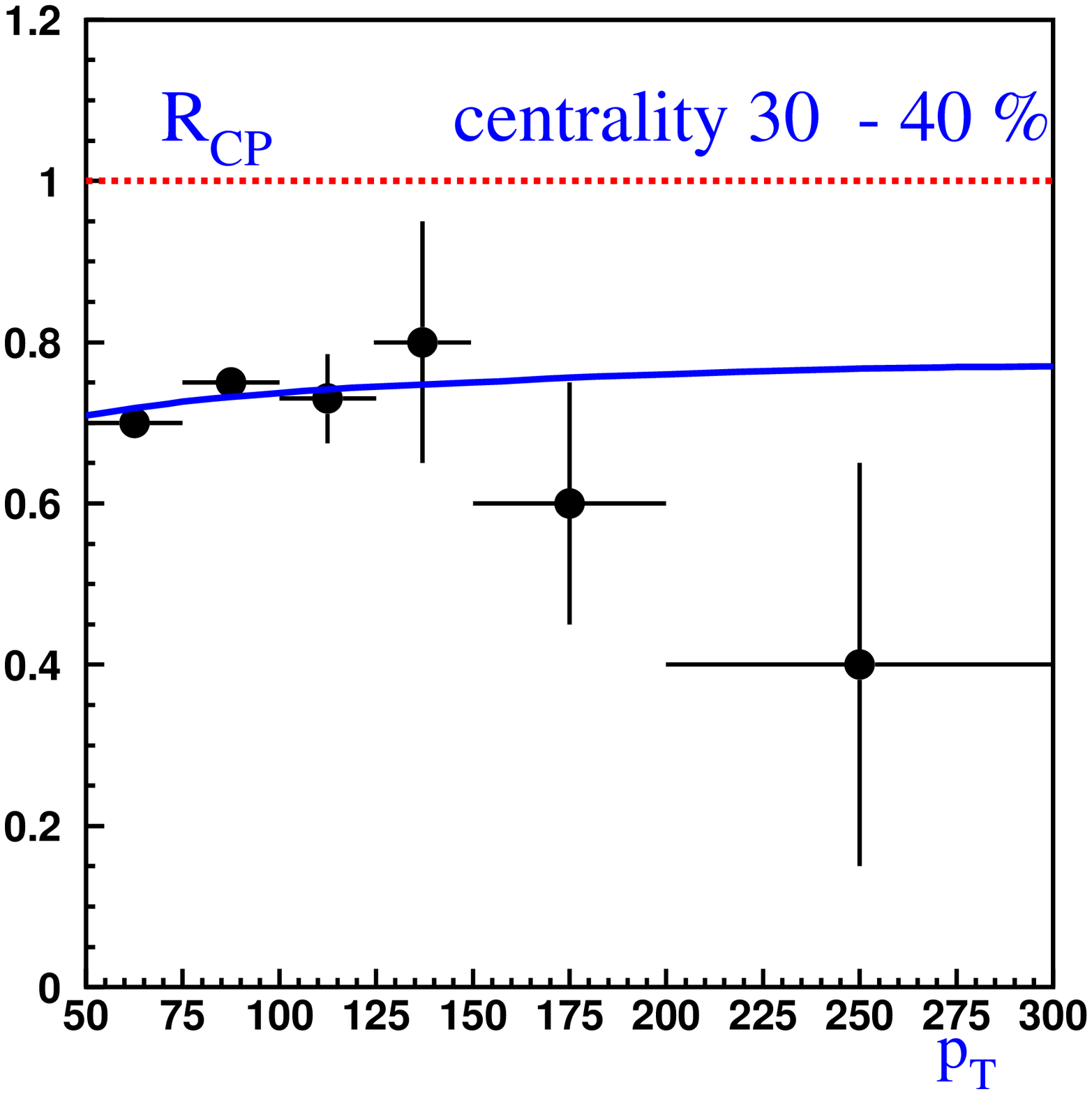} &\includegraphics[width=5cm]{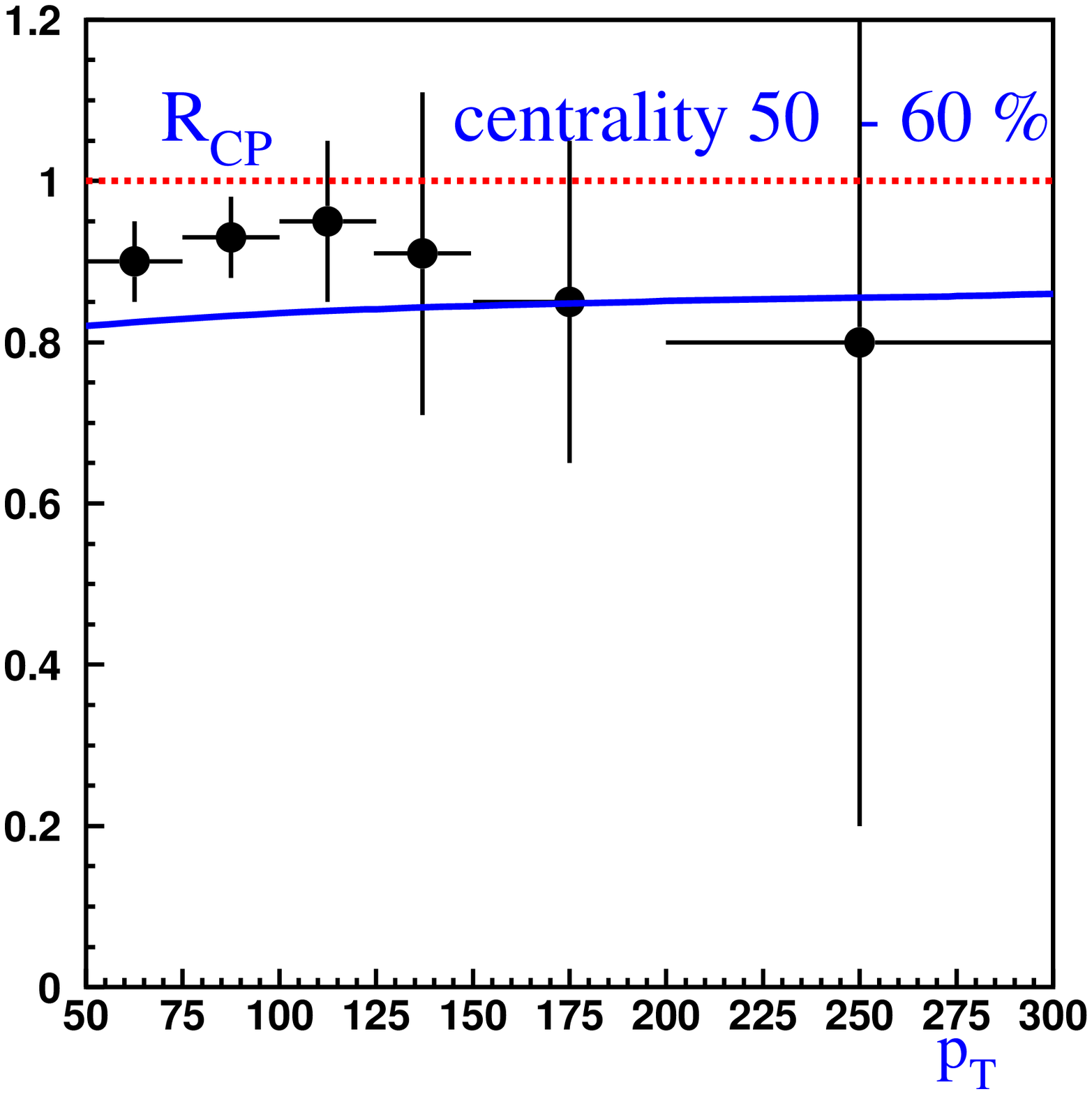}\\
\end{tabular}
\caption{Nuclear modification factor (NMF) for jet production.  The data is taken from Ref.\cite{ATLAS}. The  curves
 are calculated from \eq{RCP} while the value of the saturation momenta for each centrality events are estimated in the KLN-procedure \cite{KLN},
 with the number of participants taken from Ref.\cite{ATLAS}. The peripheral collisions are  the events with centrality $ 80\div100\%$. The values of $x_1$ and $x_2$ are chosen to be $x_1 = x_2 = 10^{-3}$.
}
\label{rcp}
\end{figure}

Using the KLN approach \cite{KLN}, we can estimate  $R_{CP}$ which is defined as 
\bea \label{RCP}
R_{CP}\,\,&=&\,\,\frac{1}{N^{centr}_{coll}}\frac{d N^{centr}}{d Y d^2 p_T}{\Big/}\frac{1}{N^{periph}_{coll}}\frac{d N^{periph}}{d Y d^2 p_T}\,\,\\
&=&\,\,\frac{\frac{d}{d p^2_T}\Big( x_1 G_P\Lb x_1, p^2_T/Q^2_s\Lb centr; x_{10}\Rb \Rb\, x_2 G_P\Lb x_2, p^2_T/Q^2_s\Lb centr; x_{20}\Rb\Rb\Big)}{\frac{d}{d p^2_T}\Big(x_1 G_P\Lb x_1, p^2_T/Q^2_s\Lb periph; x_{10}\Rb \Rb\, x_2 G_P\Lb x_2, p^2_T/Q^2_s\Lb periph; x_{20}\Rb\Rb\Big)}\nn
\eea

where $N^{centr}_{coll}$($Q \Lb centr; Y_0\Rb$) and $N^{periph}_{coll}$($Q \Lb periph; Y_0\Rb$)  are the number of collisions  ( saturation momentum) in  the events with fixed centrality and for peripheral collision, respectively.
  Based on the KLN approach we know the saturation momentum for the events with fixed centrality, and using \eq{RCP} we calculate the $R_{CP}$ (see \fig{rcp}).

\section{Conclusions}

The main result of the paper is  that  the two questions which  have been formulated in the introduction, namely,
could we obtain the NMF less that unity for production of gluon  at very high energy (small $x$)  and with large value of transverse momentum, without violation of the factorization theorem; and  can be the value of NMF of the order of the value measured by ATLAS,  have the affirmative answers.

 \eq{AGF} shows that the NMF for the gluon jet production turns out to be less than 1.
 This equation is proven using two main ingredients from the Colour Glass Condensate (CGC)/saturation approach:  the typical $x_0$ in the initial condition for the evolution in the region of low $x$ even at high values of $p_T$  is rather low and we need to use the Color Glass Condensate approach to describe them; and   the initial condition at fixed $Y=Y_0$ depends on  one variable
 $\tau=p^2_T/Q^2_s(Y_0)$, and this initial condition has the same form for  the scattering with nuclei and protons. The last assumption is essential for the estimates
 of $R_{AA}$, but  
not for $R_{CP}$. For $R_{CP}$ we can use a weaker assumption, namely,  that the McLerran-Venugopalan formula  can be used for the description of the peripheral collisions.

As have been claimed in the introduction we do not pretend that we can describe the data since we consider the theoretical example of very small  and fixed $x$  and large values of $p_T$ .  The kinematical region were the experimental data are taken, perhaps, is quite different. Our estimates show  that the scale of the effect is of the oder of the experimental one.
Therefore,  the fact that we do not reach a good agreement with the experimental  data
does not discourage us. It should be mentioned that  the simple formula that we use (see \eq{GFIN})
 relies on the saddle point approximation, and on the  particularly simple form of \eq{MVF} for the McLerran-Venugopalan formula.
 
 For serious comparison with the experimental data we have to develop an approach similar to one in Ref.\cite{ALMAR}.
 We believe that we find the explanation, why NMF calculated in this paper as well is in many others based on CGC approach , lead to the value of the NMF less than unity.

 The conclusion from the paper can be formulated in one sentence: the NMF can be smaller that unity for jet production at low and fixed  $x$ and  at large transverse  momentum $p_T$ without any violation of the factorization theorem and the initial state effects are able to explain the NMF of the order of the one measured by ATLAS. In other word, initial state effects are able to describe the jet quenching for the gluon jet production.
 
 This mechanism leads to stronger suppression at smaller value of $p_T$  (see \fig{raa}-b) and it should be taken into
account in the explanation of the NMF for produced hadrons. 

\section{Acknowledgements}

We thank Boris Kopeliovich for fruitful discussions on the subject, that convinced  me that the origin of the NMF is deeper than the interaction of the jet
 in the final state.
This research was supported by the Fondecyt (Chile) grant 1100648. 

%
\end{document}